\documentclass[12pt,english]{article}
\usepackage[T1]{fontenc}
\usepackage[latin1]{inputenc}
\usepackage{a4wide}
\usepackage{graphicx}
\usepackage{amssymb}

\makeatletter

\usepackage{babel}
\makeatother
\begin{document}

\title{Noise Characteristics of Feed Forward Loops}

\author{Bhaswar Ghosh, Rajesh Karmakar and Indrani Bose{*}}

\maketitle
\begin{center}Department of Physics\\
Bose Institute\\
93/1, A. P. C. Road\\
Kolkata - 700 009, India\end{center}

{*}Author to be contacted for correspondence; e-mail: indrani@bosemain.boseinst.ac.in

\begin{abstract}
A prominent feature of gene transcription regulatory networks is the
presence in large numbers of motifs, i.e, patterns of interconnection,
in the networks. One such motif is the feed forward loop (FFL) consisting
of three genes $X$, $Y$ and $Z$. The protein product of $x$ of
$X$ controls the synthesis of protein product $y$ of $Y$. Proteins
$x$ and $y$ jointly regulate the synthesis of $z$ proteins from
the gene $Z$. The FFLs, depending on the nature of the regulating
interactions, can be of eight different types which can again be classified
into two categories: coherent and incoherent. In this paper, we study
the noise characteristics of FFLs using the Langevin formalism and
the Monte Carlo simulation technique based on the Gillespie algorithm.
We calculate the variances around the mean protein levels in the steady
states of the FFLs and find that, in the case of coherent FFLs, the
most abundant FFL, namely, the Type-1 coherent FFL, is the least noisy.
This is however not so in the case of incoherent FFLs. The results
suggest possible relationships between noise, functionality and abundance.

\vspace{0.5cm}

\end{abstract}

\section*{1. Introduction}

Biological networks represent the complex webs of biomolecular interactions
and reactions underlying cellular processes. Well-known examples of
biological networks include metabolic reaction, protein-protein interaction
and gene transcription regulatory networks (GTRNs) \cite{key-1,key-2}.
The availability of large scale experimental data and powerful computational
tools provide information on the structural and functional features
of the complex networks. In the case of a GTRN, the nodes of the network
represent genes and two nodes are connected by a directed link if
the protein product of one gene regulates the synthesis of proteins
from the other gene. Existing databases on simple organisms like \emph{E.
coli} and \emph{S. cerevisiae} show that the GTRNs of these organisms
have common structural motifs like bi-fan, single input module (SIM)
and feed forward loop (FFL) \cite{key-3,key-4,key-5}. Such motifs
are more abundant in the naturally occurring networks than in their
randomized counterparts, highlighting the essential roles of motifs
in network function. 

The regulatory and other biochemical processes associated with a GTRN
are probabilistic in nature giving rise to fluctuations in the levels
of proteins synthesized by different genes. The magnitude of noise
cannot be neglected when the number of biomolecules participating
in the network processes is small. Recently, several theoretical \cite{key-6,key-7,key-8,key-9,key-10}
as well as experimental \cite{key-11,key-12,key-13} studies have
been carried out on the origins and consequences of stochasticity
and the dependence of noise on some important parameters of gene expression
(GE) like the transcription and translation rates. The effect of stochasticity
may be both advantageous and disadvantageous. Stochasticity can give
rise to phenotypic variations in an identical population of cells
kept in the same environment. It thus plays a  positive role in situations
where phenotypic diversity is beneficial. In most cases, however,
stochasticity acts to diminish fidelity in cellular processes. Noisy
regulatory signals, for example, may not achieve the desired outcome
introducing uncertainty in cellular behaviour.

\begin{figure}
\begin{center}\includegraphics[%
  width=2.5in]{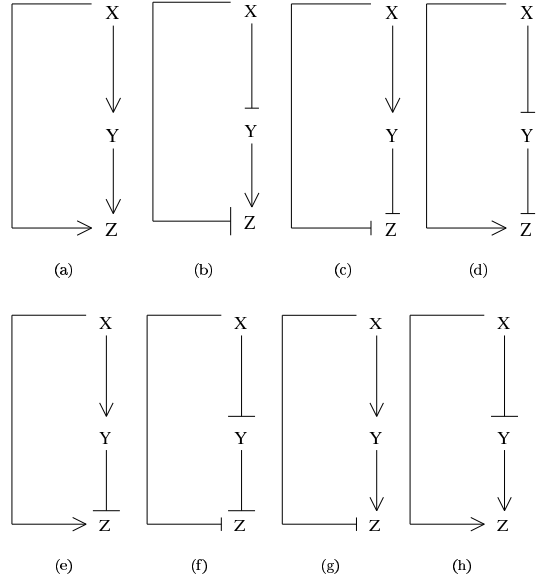}\end{center}

\caption{Eight types of FFLs: (a) Type-1, (b) Type-2, (c) Type-3, (d) Type-4 coherent FFLs, (e) Type-1, (f) Type-2, (g) Type-3, (h)Type-4 incoherent FFLs. The arrow sign denotes activation and the $\perp$ sign repression.   }
\end{figure}

Fraser et al. \cite{key-14} have recently addressed the important
issue of the relation of noise to the fitness of an organism. They
estimate the noise in protein production for almost all the genes
in \emph{S. cerevisiae} and show that the amount of noise associated
with protein levels in the steady state has lower magnitude in the
cases of essential genes and genes encoding subunits of multi-protein
complexes. Fluctuations in the protein levels of these functionally
important classes of genes are particularly detrimental to organismal
fitness because of reduced functionality. The lower amounts of noise
associated with the genes support the hypothesis that noise is an
evolvable trait acted on by natural selection. In this paper, we consider
a simple stochastic model of GE to determine the noise characteristics
of a particular type of motif appearing in GTRNs, namely, the FFL
\cite{key-3,key-4,key-5}. A FFL is a three-node motif describing
three genes $X$, $Y$ and $Z$ (figure 1). The protein $x$ produced
from gene $X$ regulates protein synthesis from gene $Y$. Proteins
$x$ and $y$ also jointly regulate the expression of gene $Z$. Inducer
molecules $S_{x}$ and $S_{y}$ are in general required to activate
or inhibit the function of protein molecules $x$ and $y$. There
are three transcriptional regulatory interactions in a FFL, each of
which can have either positive (activation) or negative (repression)
sign. The motif with three links can be in eight possible configurations
which fall into two categories: coherent and incoherent (figure 1).
In a coherent FFL, the sign of the direct regulation path from $X$
to $Z$ is the same as the overall sign of the indirect regulation
path via $Y$. There are four such configurations. In the other four
configurations, termed incoherent FFLs, the signs of the direct and
indirect regulation paths are opposite. The two protein inputs $x$
and $y$ regulate the target gene $Z$ through either an AND-gate
or an OR-gate. In the first case, both $x$ and $y$ proteins are
needed to regulate gene $Z$ and in the second case, either $x$ or
$y$ protein is sufficient for the regulation of $Z$. The functionality
of the different types of FFLs has been determined using a simple
mathematical analysis based on the deterministic rate equation approach
\cite{key-4}. The coherent FFL is found to serve as a sign-sensitive
delay element. Consider the Type-1 coherent FFL with AND-gate regulation
and a step-like pulse of $x$ proteins as the input stimulus (signal).
Expression of gene $Z$ can only begin when the level of $y$ proteins
is sufficient to cross the activation threshold for $Z$. The response
time is a measure of the speed of response and is given by the time
taken for the $z$ proteins to reach an amount which is half the steady
state level. Sign-sensitive delay implies that the response time to
step-like stimuli is asymmetric, i.e, the response time is delayed
in one direction (pulse OFF to ON) and rapid in the other direction
(ON to OFF). As a result, if the activation of the $X$ gene is transient,
the $Z$ gene cannot be significantly activated, i.e, the input signal
is not transduced through the FFL. The $z$ proteins are synthesized
only when the $X$ gene is activated for a sufficiently long time
interval. The $Z$ gene switches off rapidly once the $X$ gene is
deactivated. In other words, the coherent FFL functions as a persistence
detector, responding only to a persistent stimulus and filtering out
fluctuations in the input signal. The role of the coherent FFL as
sign-sensitive delay has been verified experimentally \cite{key-15}.
The incoherent FFLs function as sign-sensitive accelerators speeding
up the response time in one direction (OFF to ON in the stimulus step)
but not in the other direction (ON to OFF). Some incoherent FFLs act
also as pulse generators. Amongst the coherent FFLs, the Type-1 FFL
appears the maximum number of times in the GTRNs of \emph{E. coli}
and \emph{S. cerevisiae.} Similarly, in the case of incoherent FFLs,
the Type-1 FFL is the most abundant. We calculate the noise characteristics
of the coherent and incoherent FFLs using the Langevin formalism \cite{key-16}
and the Monte Carlo simulation technique based on the Gillespie algorithm
(GA) \cite{key-17,key-18}. We show that the most abundant coherent
FFL, namely, the Type-1 FFL, is the least noisy. This is, however,
not true in the case of the incoherent FFLs. The lower number of FFLs
has been ascribed to their reduced functionality \cite{key-4}. Noise
is disadvantageous if it affects operational reliability. Our results
on noise characteristics of FFLs suggest that noisy motifs are likely
to be selected against during evolution if noise is detrimental to
the function of the motifs.

\section*{2. Stochastic Model of GE}

The simple stochastic model of GE has been studied earlier as a Markovian
model for the gene induction process \cite{key-19} and also to explore
the possible origins of the genetic disorder, haploinsufficiency \cite{key-10,key-20}.
In the minimal model, a gene can be in two possible states: inactive
($G$) and active ($G^{*}$). Due to stochasticity, the gene makes
random transitions between the inactive and active states with $k_{a}$
and $k_{d}$ being the activation and deactivation rate constants.
In the active state, protein production occurs with the rate constant
$\beta_{p}$. Protein decay occurs with the rate constant $\gamma_{p}$.
The protein decay rate has two components, one, the degradation rate
and the other, the dilution rate of proteins due to cell growth and
division. The reaction scheme RS-1 is shown in equation (\ref{mathed:First equation}),\begin{equation}
G\quad\begin{array}{c}
k_{a}\\
\rightleftharpoons\\
k_{d}\end{array}\quad G^{\star}\quad\begin{array}[b]{c}
\beta_{p}\\
\longrightarrow\end{array}\quad p\quad\begin{array}[b]{c}
\gamma_{p}\\
\longrightarrow\end{array}\quad\Phi\label{mathed:First equation}\end{equation}

Let $P(n_{1},n_{2},t)$ be the probability that at time $t$, $n_{1}$
genes are in the active state $G^{*}$ and the number of protein molecules
is $n_{2}$. The rate of change of the probability with respect to
time is given by the Master Equation\begin{equation}
\begin{array}{cc}
\frac{\partial P(n_{1},n_{2},t)}{\partial t}= & k_{a}[(n_{tot}-n_{1}+1)P(n_{1}-1,n_{2},t)-(n_{tot}-n_{1})P(n_{1},n_{2},t)]\\
 & +k_{d}[(n_{1}+1)P(n_{1}+1,n_{2},t)-n_{1}P(n_{1},n_{2},t)]\\
 & +\beta_{p}[n_{1}P(n_{1},n_{2}-1,t)-n_{1}P(n_{1},n_{2},t)]\\
 & +\gamma_{p}[(n_{2}+1)P(n_{1},n_{2}+1,t)-n_{2}P(n_{1},n_{2},t)]\end{array}\label{mathed:Second Equation}\end{equation}

\noindent where $n_{tot}$ is the total number of genes.

For each rate constant, the gain term adds to the probability and
the loss term subtracts from the same. The simplicity of the stochastic
model enables one to calculate the mean protein level $<n_{2}>$ and
its variance $<\delta n_{2}^{2}>=<n_{2}^{2}>-<n_{2}>^{2}$ in the
steady state using the standard generating function approach. The
results are:\begin{equation}
<n_{2}>=\frac{\beta_{p}}{\gamma_{p}}\,\frac{n_{tot}\, k_{a}}{k_{a}+k_{d}}\label{mathed: Third Equation}\end{equation}
\begin{equation}
<\delta n_{2}^{2}>=<n_{2}>[1+\frac{\beta_{p}\, k_{d}}{(k_{a}+k_{d})(k_{a}+k_{d}+\gamma_{p})}]\label{mathed: Fourth Equation}\end{equation}

\noindent Also, the mean number of genes in the active state is given
by \begin{equation}
<n_{1}>=\frac{n_{tot}\, k_{a}}{k_{a}+k_{d}}\label{mathed: Fifth Equation}\end{equation}

\noindent The minimal model (equation (\ref{mathed:First equation}))
describes constitutive GE. We now assume that the transition from
the state $G$ to the state $G^{*}$ is brought about by activating
regulatory molecules $S.$ The reaction scheme RS-2 in the presence
of such molecules is given by\begin{equation}
G+S\quad\begin{array}{c}
k_{1}\\
\rightleftharpoons\\
k_{2}\end{array}\quad G_{-}S\quad\begin{array}{c}
k_{a}\\
\rightleftharpoons\\
k_{d}\end{array}\quad G^{\star}\quad\begin{array}[b]{c}
\beta_{p}\\
\longrightarrow\end{array}\quad p\quad\begin{array}[b]{c}
\gamma_{p}\\
\longrightarrow\end{array}\quad\Phi\label{mathed: Sixth Equation}\end{equation}

\noindent where $G_{-}S$ represents the bound complex of $G$ and
$S$ from which transition to the active state $G^{*}$ occurs. The
total number of genes $n_{tot}$ is given by\begin{equation}
n_{tot}=g+g_{s}+g^{*}\label{mathed: Seventh Equation}\end{equation}

\noindent  where $g,$ $g_{s}$ and $g^{*}$ are the number of genes
in the states $G,$ $G_{-}S$ and $G^{*}$ respectively. In the steady
state, $\frac{dg}{dt}=0$ and $\frac{dg^{*}}{dt}=0$. From the first
condition, one obtains \begin{equation}
\frac{g\, s}{K_{1}}=g_{s}\label{mathed: Eighth equation}\end{equation}

\noindent where $K_{1}=\frac{k_{2}}{k_{1}}$ is the equilibrium dissociation
constant and $s$ is the number of regulatory molecules. From the
second condition, the expression for $g^{*}$ in the steady state
is given by\begin{equation}
g^{*}=\frac{n_{tot}k_{a}\frac{s/K_{1}}{1+s/K_{1}}}{k_{a}\frac{s/K_{1}}{1+s/K_{1}}+k_{d}}\label{mathed: 9}\end{equation}

\noindent  Expressions (\ref{mathed: Fifth Equation}) and (\ref{mathed: 9})
for the number of genes in the active state $G^{*}$ are equivalent
on defining effective activation  and deactivation rate constants

\begin{equation}
k_{a}^{'}=k_{a}\:\frac{s/K_{1}}{1+s/K_{1}}\quad\quad k_{d}^{'}=k_{d}\label{mathed: 10}\end{equation}

\noindent  The equivalence relations are useful as one can map the
reaction scheme RS-2 onto the simpler scheme RS-1 while calculating
mean protein levels and the associated variances. Regulatory molecules,
in general, oligomerise to form an active complex $S_{n}$ where $n$
is the number of regulatory molecules contained in the complex. In
this case, the effective rate constants $k_{a}^{'}$ and $k_{d}^{'}$
are given by

\begin{equation}
k_{a}^{'}=k_{a}\frac{(s/K)^{n}}{1+(s/K)^{n}},\quad\quad k_{d}^{'}=k_{d}\label{mathed: 11}\end{equation}

\noindent  where $K^{n}=K_{1}K_{c}$ , $K_{c}$ being the equilibrium
dissociation constant for oligomerisation, \hspace{3mm}i. e, 

\begin{equation}
n\, S\quad\begin{array}[b]{c}
K_{c}\\
\rightleftharpoons\end{array}\quad S_{n}\label{mathed: 12}\end{equation}

\noindent  When the regulatory molecules $S$ act as repressors,
the effective rate constants are given by

\begin{equation}
k_{a}^{'}=k_{a}\frac{1}{1+(s/K)^{n}},\quad\quad k_{d}^{'}=k_{d}\label{mathed: 13}\end{equation}

\noindent  In this case, repressor molecules on binding to genes
prevent their activation to the state $G^{*}$.

We now apply the stochastic model of GE to determine the mean levels
of proteins $x$, $y$ and $z$ and the variances thereof in the steady
state of a FFL. The variances calculated are a measure of the intrinsic
noise associated with GE as fluctuations in the number of regulatory
molecules are ignored. Let $\beta_{i}$ and $\gamma_{i}$ ($i=x,\,\, y,\,\, z)$
be the rate constants for the synthesis and decay respectively of
protein $i$. For proteins $x$, the mean protein level $x_{av}$
and its variance $<\delta x^{2}>$ in the steady state are obtained
from equations (\ref{mathed: Third Equation}) and (\ref{mathed: Fourth Equation})
\cite{key-10,key-19} as (with $n_{tot}=1$)

\begin{equation}
x_{av}=<x>=\frac{\beta_{x}}{\gamma_{x}}\:\:\frac{k_{a}}{k_{a}+k_{d}}\label{mathed: 14}\end{equation}

\begin{equation}
<\delta x^{2}>=<x>\:\:[1+\frac{\beta_{x\,\,}k_{d}}{(k_{a}+k_{d})(k_{a}+k_{d}+\gamma_{x})}]\label{mathed:15}\end{equation}

\noindent where $k_{a}$ and $k_{d}$ are activation and deactivation
rate constants of gene $X$. Protein molecules $x$ regulate the activation
of gene $Y$ according to the reaction scheme RS-2. Mapping onto the
simpler reaction scheme RS-1, one obtains in the steady state

\begin{equation}
y_{av}=<y>=\frac{\beta_{y}}{\gamma_{y}}\:\:\frac{k_{a}^{'}}{k_{a}^{'}+k_{d}^{'}}\label{mathed: 16}\end{equation}

\begin{equation}
<\delta y^{2}>=<y>\:\:[1+\frac{\beta_{y\,\,}k_{d}^{'}}{(k_{a}^{'}+k_{d}^{'})(k_{a}^{'}+k_{d}^{'}+\gamma_{y})}]\label{mathed: 17}\end{equation}

\noindent The effective rate constants $k_{a}^{'}$ and $k_{d}^{'}$
have the forms given in equations (\ref{mathed: 11}) or (\ref{mathed: 13})
depending on whether the regulatory interaction is activating or repressing
in nature. In the case of activation, assuming $n$ to be $2$, 

\begin{equation}
k_{a}^{'}=k_{ay}\,\,\frac{(x/K_{xy})^{2}}{1+(x/K_{xy})^{2}},\quad\quad k_{d}^{'}=k_{dy}\label{mathed: 18}\end{equation}

\noindent In (\ref{mathed: 18}), $k_{ay}$ represents the limiting
value of $k_{a}^{'}$ obtained when $\frac{x}{K_{xy}}\gg1$. In the
case of repression, 

\begin{equation}
k_{a}^{'}=k_{ay}\,\,\frac{1}{1+(x/K_{xy})^{2}},\quad\quad k_{d}^{'}=k_{dy}\label{mathed: 19}\end{equation}

\noindent Both the $x$ and $y$ proteins regulate the activation
of the $Z$ gene. The mapping of the associated reaction scheme onto
the simpler reaction scheme RS-1 is still possible. The effective
rate constants $k_{a}^{''}$ and $k_{d}^{''}$ have specific forms
depending on the nature of the regulating interaction (activating/
repressing) and the type of logic gate (AND/ OR) in operation. The
mean protein level in the steady state and its variance are\begin{equation}
z_{av}=<z>=\frac{\beta_{z}}{\gamma_{z}}\:\:\frac{k_{a}^{''}}{k_{a}^{''}+k_{d}^{''}}\label{mathed: 20}\end{equation}

\begin{equation}
<\delta z^{2}>=<z>\:\:[1+\frac{\beta_{z\,\,}k_{d}^{''}}{(k_{a}^{''}+k_{d}^{''})(k_{a}^{''}+k_{d}^{''}+\gamma_{z})}]\label{mathed:21}\end{equation}

\noindent The activation and deactivation rate constants $k_{a}^{''}$
and $k_{d}^{''}$ are

\begin{equation}
k_{a}^{''}=k_{az}\,\, G(x,y,T_{xz}\,,T_{yz}),\quad\quad k_{d}^{''}=k_{dz}\label{mathed: 22}\end{equation}

\noindent where $k_{az}$ is the limiting value of $k_{a}^{''}$.
For the AND-gate,

\begin{equation}
G(x,y,T_{xz}\,\,,T_{yz})=T_{xz}\:\: T_{yz}\label{mathed:23}\end{equation}

\noindent For the OR-gate,

\begin{equation}
G(x,y,T_{xz}\,\,,T_{yz})=\frac{(1+(x/K_{xz})^{2})\, T_{xz}+(1+(y/K_{yz})^{2})\, T_{yz}}{1+(x/K_{xz})^{2}+(y/K_{yz})^{2}}\label{mathed:24}\end{equation}

\noindent This expression has been derived assuming that the regulatory
molecules $x$ and $y$ compete to bind at the operator region of
the gene $Z$, as in Ref. \cite{key-4}.

\noindent For activating regulatory interactions,

\begin{equation}
T_{xz}=\frac{(x/K_{xz})^{2}}{1+(x/K_{xz})^{2}}\,\,,\quad\quad T_{yz}=\frac{(y/K_{yz})^{2}}{1+(y/K_{yz})^{2}}\label{mathed:25}\end{equation}

\noindent For repressing regulatory interactions,

\begin{equation}
T_{xz}=\frac{1}{1+(x/K_{xz})^{2}}\,\,,\quad\quad T_{yz}=\frac{1}{1+(y/K_{yz})^{2}}\label{mathed:26}\end{equation}

\noindent The parameters $K_{xy}$, $K_{yz}$ and $K_{xz}$ appearing
in (\ref{mathed: 17}), (\ref{mathed:25}) and (\ref{mathed:26})
are analogous to the parameter $K$ in (\ref{mathed: 11}). In the
steady state of the FFL, all three proteins $x$, $y,$ $z$ are in
their steady state levels and the effective rate constants $k_{a}^{'}$,
$k_{a}^{''}$ are calculated with the steady state values $x=x_{av}$
and $y=y_{av}$.

The FFL may be considered to be a two-step signaling cascade. The
$x$ and $z$ proteins constitute respectively the input and output
signals of the cascade. With stochasticity taken into account, it
is desirable that cascades are able to transmit signals in a reliable
manner. When fluctuations are considerable, there is a danger of the
noise building up in successive steps of the cascade corrupting the
final output signal. Thattai and Oudenaarden \cite{key-16} have studied
the noise characteristics of signaling cascades and have shown that
under certain conditions the fluctuations in the output signal are
bounded. Also, noise reduction is possible, i.e, the output signal
is less noisy than the input signal.

\section*{3. Noise Characteristics of FFL}

The variance around the mean protein level has two components: intrinsic
and extrinsic. In the last section, the variance due to only the intrinsic
part has been calculated. In this section, the fluctuations in the
number of regulatory molecules, constituting extrinsic noise, are
taken into account. The total variances in the steady state of the
FFL are denoted as $<\delta x^{2}>_{tot}$ (equals $<\delta x^{2}>$
given in $(15)$), $<\delta y^{2}>_{tot}$ and $<\delta z^{2}>_{tot}$.
The variances can be calculated using the method followed in \cite{key-16}.
We use Langevin equations to take stochasticity into account. The
equation describing the production of protein $x$ is given by

\begin{equation}
\dot{x}=\beta_{x}\,\,\frac{k_{a}}{k_{a}+k_{d}}-\gamma_{x}\,\, x+\eta_{1}(t)\label{mathed: 27}\end{equation}

\noindent  where $\dot{x}$ represents a time derivative. Stochasticity
is associated with the time-dependent noise term $\eta_{1}(t)$ in
equation $(27)$. The random variable $\eta_{1}(t)$ obeys white-noise
statistics, i.e, 

\begin{equation}
<\eta_{1}(t)>=0,\quad\quad<\eta_{1}(t)\,\eta_{1}(t+\tau)>=q_{1}\,\delta(\tau)\label{mathed: 28}\end{equation}

\noindent where $\delta(\tau)$ is the Dirac delta function and $<...>$
denotes an ensemble average. The state dependences of $\eta_{1}(x,t)$
and $q(x)$ are ignored since we are interested in the steady state
noise characteristics. In the absence of the noise term in equation
(\ref{mathed: 27}), the mean protein level in the steady state ($\dot{x}(t)=0)$,
as in equation (\ref{mathed: 14}), is recovered. We linearize equation
(\ref{mathed: 27}) for fluctuations, assumed to be small, about the
steady state to obtain 

\begin{equation}
\delta\dot{x}(t)+\gamma_{x}\,\delta x=\eta_{1}(t)\label{mathed: 29}\end{equation}

\noindent Fourier transform of equation $(29)$ yields

\begin{equation}
(i\,\omega+\gamma_{x})\,\,\delta x(\omega)=\eta_{1}(\omega)\label{mathed: 30}\end{equation}

\noindent Next, taking ensemble average and applying condition $(28)$,
we get

\begin{equation}
<|\delta x(\omega)|^{2}>=\frac{q_{1}}{\omega^{2}+\gamma_{x}^{2}}\label{mathed:31}\end{equation}

\noindent The steady state variance $<\delta x^{2}>_{tot}$ is given
by an inverse Fourier transform at $\tau=0$, i.e, 

\begin{equation}
<\delta x^{2}>_{tot}=\frac{q_{1}}{2\,\gamma_{x}}\label{mathed: 32}\end{equation}

\noindent Since $<\delta x^{2}>_{tot}=<\delta x^{2}>$ (equation
(\ref{mathed:15})), $q_{1}$ is known explicitly from equation (\ref{mathed: 32}).
For protein $y$, the Langevin equation is given by

\begin{equation}
\dot{y}+\gamma_{y}\, y=\beta_{y}\, f_{xy}(x)+\eta_{2}(t)\label{mathed: 33}\end{equation}

\noindent with 

\begin{equation}
<\eta_{2}(t)\,\eta_{2}(t+\tau)>=q_{2}\,\delta(\tau)\label{mathed: 34}\end{equation}

\noindent In equation (\ref{mathed: 33}), the rate of creation of
$y$ proteins in terms of the $x$ proteins is given by the first
term on the r.h.s. The function $f_{xy}(x)$ is designated as the
transfer function and is given by

\begin{equation}
f_{xy}(x)=\frac{k_{a}^{'}}{k_{a}^{'}+k_{d}^{'}}\label{mathed: 35}\end{equation}

\noindent where $k_{a}^{'}$ and $k_{d}^{'}$ are as defined in equations
(\ref{mathed: 18}) and (\ref{mathed: 19}). Again, the mean protein
level in the steady state, $y_{av}$ (equation (\ref{mathed: 16}))
can be recovered from equation (\ref{mathed: 33}) by ignoring the
noise term and putting $\dot{y}=0$. Going through the same steps
as before, the variance $<\delta y^{2}>_{tot}$ is obtained as

\begin{equation}
<\delta y^{2}>_{tot}=\frac{q_{2}}{2\,\gamma_{y}}+\frac{\beta_{y}^{2}\, c_{x}^{2}\, q_{1}}{2\,\gamma_{x}\,\gamma_{y}\,(\gamma_{x}+\gamma_{y})}\label{mathed:36}\end{equation}

\noindent The first term in equation $(36)$ is the intrinsic noise
term given by $<\delta y^{2}>$ (equation (\ref{mathed: 17})). The
second term, describing extrinsic noise, arises due to the noise propagated
from the input, i.e, due to the fluctuations in the number of $x$
regulatory proteins. In the same equation, $c_{x}$ is the derivative
of the transfer function $f_{xy}(x)$, w.r.t $x$, calculated at the
steady state value of $x$, i.e 

\begin{equation}c_{x}=\frac{{\partial}{f_{xy}(x)}}{{\partial}{x}}\left |\begin{array}{c} \\x=x_{av} \end{array} \right . \label{mathed:37}\end{equation} 

\noindent For the $z$ proteins, the Langevin equation is

\begin{equation}
\dot{z}+\gamma_{z}\, z=\beta_{z}\, g_{xy}(x,y)+\eta_{3}(t)\label{mathed:38}\end{equation}

\noindent with 

\begin{equation}
<\eta_{3}(t)\,\eta_{3}(t+\tau)>=q_{3}\,\delta(\tau)\label{mathed:39}\end{equation}

\noindent The transfer function $g_{xy}(x,y)$ is given by

\begin{equation}
g_{xy}(x,y)=\frac{k_{a}^{''}}{k_{a}^{''}+k_{d}^{''}}\label{mathed:40}\end{equation}

\noindent where $k_{a}^{''}$ and $k_{d}^{''}$ have been defined
in equations (\ref{mathed: 22})-(\ref{mathed:26}). The variance
$<\delta z^{2}>_{tot}$ is given by

\begin{equation}
\begin{array}{cc}
<\delta z^{2}>_{tot}= & \frac{q_{3}}{2\,\gamma_{z}}+\frac{q_{2}\,\beta_{z}^{2}\, d_{y}^{2}}{2\,\gamma_{y}\,\gamma_{z}\,(\gamma_{y}+\gamma_{z})}+\frac{q_{1}\,\beta_{z}^{2}\, d_{x}^{2}}{2\,\gamma_{x}\,\gamma_{z}\,(\gamma_{x}+\gamma_{z})}+\frac{q_{1}\,\beta_{y}^{2}\,\beta_{z}^{2}\, c_{x}^{2}\, d_{y}^{2}\,(\gamma_{x}+\gamma_{y}+\gamma_{z})}{2\,\gamma_{x}\,\gamma_{y}\,\gamma_{z}\,(\gamma_{x}+\gamma_{y})(\gamma_{y}+\gamma_{z})(\gamma_{x}+\gamma_{z})}\\
\\ & +\frac{q_{1}\,\beta_{y}\,\beta_{z}^{2}\,\gamma_{y}\, c_{x}\, d_{x}\, d_{y}\,(\gamma_{x}+\gamma_{y}+\gamma_{z})}{\gamma_{x}\,\gamma_{y}\,\gamma_{z}\,(\gamma_{x}+\gamma_{y})(\gamma_{y}+\gamma_{z})(\gamma_{x}+\gamma_{z})}\end{array}\label{mathed:41}\end{equation}

\noindent where

\begin{equation}d_{x}=\frac{{\partial}{g_{xy}(x,y)}}{{\partial}{x}}\left |\begin{array}{c} \\x=x_{av}, y=y_{av}\end{array}\right.,\;\;    d_{y}=\frac{{\partial}{g_{xy}(x,y)}}{{\partial}{y}}\left |\begin{array}{c} \\x=x_{av}, y=y_{av}\end{array}\right . \label{mathed:42}\end{equation}

\noindent In equation (\ref{mathed:39}), the first term $\frac{q_{3}}{2\,\gamma_{z}}$
is the intrinsic noise term given by $<\delta z^{2}>$ (equation (\ref{mathed:21})).
The other terms represent noise propagated from the earlier stages,
i.e, occur due to fluctuations in the number of $x$ and $y$ regulatory
molecules. These terms describe the extrinsic noise.

\section*{4. Results and Discussion}

We now calculate the variances $<\delta x^{2}>_{tot}$, $<\delta y^{2}>_{tot}$
and $<\delta z^{2}>_{tot}$ for the different FFLs. Our goal is to
compare the variances for the same as well as different FFLs. For
simplicity we assume that all $\gamma_{i}$'s ($i=$$x$, $y$, $z$)=1
and $K_{xy}=K_{yz}=K_{xz}=1$. The mean levels of proteins $x$, $y$
and $z$ in the steady state are kept the same in all the cases so
that a meaningful comparison between the variances can be made. Figures
2 and 3 show the plots of $<\delta x^{2}>_{tot}$ (line with long
dashes), $<\delta y^{2}>_{tot}$ (line with short dashes) and $<\delta z^{2}>_{tot}$
(solid line) versus $\beta_{y}$ for the coherent and incoherent FFLs
respectively. The regulation of the $Z$ gene by the $x$ and $y$
proteins is achieved via the AND gate. The plots have been obtained
keeping the mean protein levels $x_{av}$, $y_{av}$ and $z_{av}$
fixed at $m=5.0.$ For this we put

\begin{equation}
\beta_{x}=\beta_{z}=10,\, k_{a}=k_{d}=20,\, k_{a}^{''}=k_{d}^{''},\, k_{az}=20,\, k_{a}^{'}=\frac{m\, k_{d}^{'}}{\beta_{y}-m},\: k_{d}^{'}=20,\label{mathed: 43}\end{equation}

\noindent For the coherent Type-1 FFL with AND-gate regulation, the
values of $k_{ay}$ and $k_{a}^{''}$ are fixed from the relations

\begin{equation}
k_{a}^{'}=k_{ay}\,\frac{m^{2}}{1+m^{2}}\label{mathed:44}\end{equation}

\noindent and

\begin{equation}
k_{a}^{''}=k_{az}\,\frac{m^{4}}{(1+m^{2})^{2}}\label{mathed: 45}\end{equation}

\noindent Equivalent relations hold true for the other types of FFLs.
An examination of figure 2 shows that the Type-1 coherent FFL is the
least noisy amongst all the coherent FFLs. The number of times the
Type-1,Type-2, Type-3 and Type-4 coherent FFLs appear in the GTRNs
of \emph{E. coli} (\emph{S. cerevisiae}) are 28, 2, 4, 1 (26, 5, 0,
0) \cite{key-4}. The most abundant coherent FFL, namely the Type-1
FFL, is the least noisy. This is not true for the incoherent FFLs.
The number of times the Type-1, Type-2, Type-3, Type-4 incoherent
FFLs appear in the GTRNs of \emph{E. coli} (\emph{S. cerevisiae})
are 5, 0, 1, 1 (21, 3, 1, 0). The most abundant incoherent FFL, namely,
the Type-1 FFL, is more noisy than, say, the Type-4 incoherent FFL,
which is practically absent in the GTRNs. 

\begin{figure}[t]
\begin{center}\includegraphics[%
  width=5in]{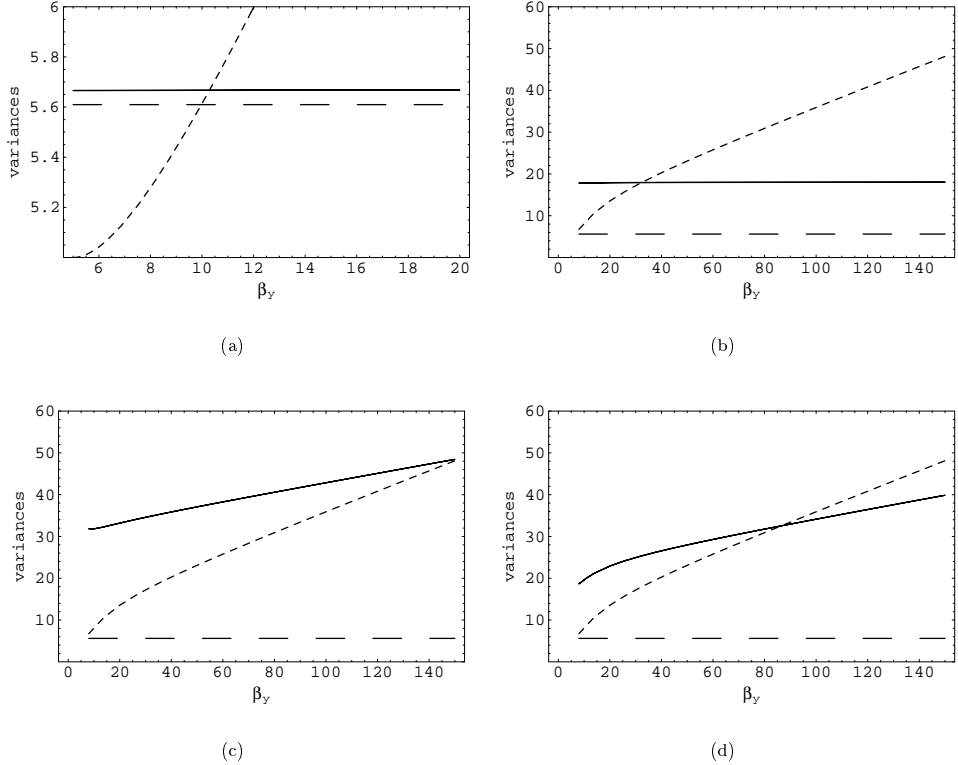}\end{center}

\caption{Variances $<\delta x^{2}>_{tot}$ (line with long dashes), $<\delta y^{2}>_{tot}$ (line with short dashes), $<\delta z^{2}>_{tot}$ (solid line) versus $\beta_{y}$ for (a) Type-1, (b) Type-2, (c) Type-3 and  (d) Type-4 coherent FFLs controlled by AND-gate. The mean protein level is fixed at m=5. The other parameter values are mentioned in the text}
\end{figure}

The reasons as to why some FFLs occur more often than the others in
GTRNs, are not well understood. Generally speaking, reduced functionality
of a motif may be a possible reason for its lower abundance, i.e,
being selected against during evolution. As suggested by Mangan and
Alon \cite{key-4}, for AND-gate FFLs, Types-3 and 4 have reduced
functionality compared to Types- 1 and 2, as the former respond to
at most one input stimulus ($S_{x}$) whereas the latter respond to
both the input stimuli $S_{x}$ and $S_{y}$. Also, Type-1 coherent
FFL gains advantage from increased cooperativity leading to a sharper
response in the presence of stimuli. For low $x$ concentrations,
the effective Hill coefficient (a measure of cooperativity) is 6 (for
$n=2$ in equation (12)) whereas the same, for the other FFLs, is
2. We now discuss the relationship between noise, function and abundance.
For the sake of clarity, we focus attention on the Type-1 and Type-4
coherent FFLs. Figure 4 shows plots for the total variances around
the mean protein level $m=5$ when the input noise $<\delta x^{2}>_{tot}$
is higher than that in the cases of figures 2 and 3. The parameter
values changed from equation (43) are $k_{a}=k_{d}=5,$ $k_{d}^{'}=30$
and $k_{az}=30$. In the case of the Type-1 coherent FFL, one finds
the existence of a parameter region in which the variance decreases
in the successive stages of the FFL so that the output noise is less
than the input noise. Such a parameter region is absent in the case
of the Type-4 coherent FFL. Another notable feature of the plots in
figures 2 and 4 is that $<\delta y^{2}>_{tot}$ and $<\delta z^{2}>_{tot}$
in the case of the Type-1 FFL have almost linear dependences on $\beta_{y}$
whereas the same quantities are more nonlinear in the case of the
Type-4 FFL. For the Type-1 FFL, the dominant contribution to $<\delta z^{2}>_{tot}$
is from the internal noise associated with the expression of the $Z$
gene. Fluctuations in the $x$ and $y$ protein levels have little
effect on the total noise. In the case of the Type-4 FFL, the extrinsic
contribution to noise is greater than that in the case of the Type-1
FFL. In short, figures 2 and 4 show that the Type-1 FFL acts as a
better filter of noise. As mentioned in the Introduction, one possible
function of coherent FFLs is as a persistent detector or equivalently
as a filter which attenuates the input noise. The Type-1 coherent
FFL being less noisy than the Type-4 coherent FFL, functions better
as a noise filter. The reduced functionality of the Type-4 coherent
FFL explains its lower abundance from an evolutionary point of view.
Similar reasoning holds true for Type-2 and Type-3 coherent FFLs.
Thus for coherent FFLs, noise is disadvantageous as it erodes the
function of a FFL as a persistent detector. For incoherent FFLs, functioning
as sign-sensitive accelerators, noise appears to have no direct relationship
with abundance, i.e, noise is not detrimental to the functioning of
the FFLs. 

\begin{figure}[t]
\begin{center}\includegraphics[%
  width=5in]{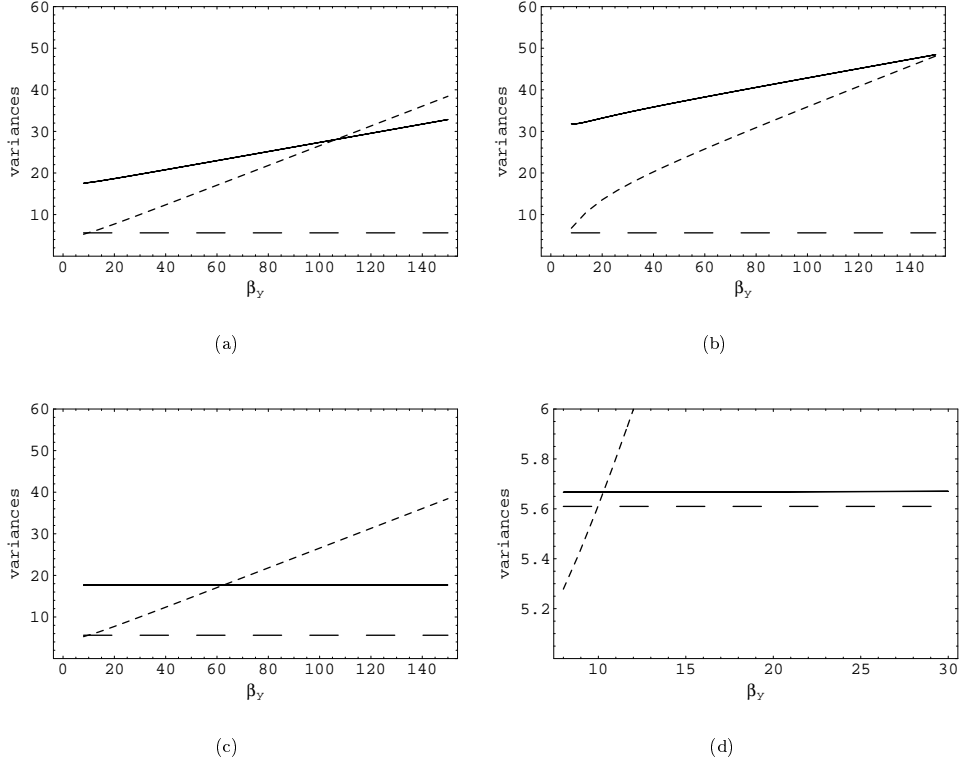}\end{center}

\caption{Variances $<\delta x^{2}>_{tot}$ (line with long dashes), $<\delta y^{2}>_{tot}$ (line with short dashes), $<\delta z^{2}>_{tot}$ (solid line) versus $\beta_{y}$ for (a) Type-1, (b) Type-2, (c) Type-3 and  (d) Type-4 incoherent FFLs controlled by AND-gate. The mean protein level is fixed at m=5. The other parameter values are mentioned in the text.}
\end{figure}

Our analysis of the noise characteristics of FFLs is based on the
Langevin formalism which is approximate in nature. To establish the
validity of the results, we have calculated the variances using Monte
Carlo simulation based on the GA \cite{key-17,key-18}. The GA provides
a numerical solution of the Master Equation leading to an accurate
description of the time course of evolution of a stochastic system.
A brief description of the GA is as follows. Consider N chemical species
participating in M chemical reactions. Let $X(i)$, $i=1,2,3,.....,N$
denotes the number of molecules of the ith chemical species. Given
the values of $X(i)$, $i=1,2,3,...N$ at time t, the GA is designed
to answer two questions: (1) when will the next reaction occur? and
(2) what type of reaction will it be? Let the next reaction occur
at time $t+\tau$. Knowing the type of reaction, one can adjust the
numbers of participating molecules in accordance with the specific
reaction scheme. Thus, with repeated applications of the GA, one can
keep track of how the numbers, $X(i)$'s, change as a function of
time due to the occurrence of $M$ different types of chemical reactions.
Each reaction $\mu$ ($\mu=1,2,3,....,M$) has a stochastic rate constants
$C_{\mu}$ associated with it. The rate constant has the interpretation
that $C_{\mu}dt$ is the probability that a particular combination
of reacting molecules participates in the $\mu th$ reaction in the
infinitesimal time interval ($t,t+dt$). If $h_{\mu}$ is the number
of distinct molecular combinations for the $\mu th$ reaction, then
$a_{\mu}dt=h_{\mu}C_{\mu}dt$ is the probability that the $\mu th$
reaction occurs in the infinitesimal time interval (t, t+dt). The
implementation of the GA algorithm is described in detail in Refs.
\cite{key-17,key-18}. We use the algorithm to determine the evolution
of the number of $z$ proteins of a FFL as a function of time. Figures
5(a) and 6(a) show the results for the coherent Type-1 and Type-4
FFL respectively. The solid line, in each case, represents the mean
trajectory obtained from a solution of the deterministic equations.
The reactions considered are those associated with a FFL. Expression
of each gene $X$ , $Y$ and $Z$ is according to the reaction scheme
RS-2 (equation (6)). For the $X$ gene, there is no regulatory molecule
$S$. The $x$ proteins dimerize (equation (12) with $n=2)$ and the
dimers regulate expression of the $Y$ gene. The $y$ proteins also
dimerize to regulate expression of the gene $Z$. Considering AND-gate
regulation of the $Z$ gene expression, both the $x$ and $y$ protein
dimers bind simultaneously at the operator region for activation of
the gene. Other possibilities like the operator region unoccupied
or occupied by a single dimer are considered but the gene remains
in the inactive state in these cases. The stochastic rate constants
$C_{\mu}$'s are equal to the rate constants $k_{\mu}$'s since in
the deterministic approach the numbers and not the concentrations
of the different molecules are considered. Figure 5(b) and 6(b) show
the histograms describing the distribution of protein levels, $N(z)$
versus $z$, for the coherent Type-1 and Type-4 FFLs respectively.
The histograms have been obtained by accumulating data over 5000 trial
runs. The distribution is broader in the case of the Type-4 coherent
FFL indicating that it is more noisy than the Type-1 FFL. The variances
for Type-1 and Type-4 distributions are 110.612 and 329.990 respectively.
The simulation results support the results obtained by using the Langevin
formalism that the Type-4 coherent FFL is more noisy than the Type-1
coherent FFL.

\begin{figure}
\begin{center}\includegraphics[%
  width=5in]{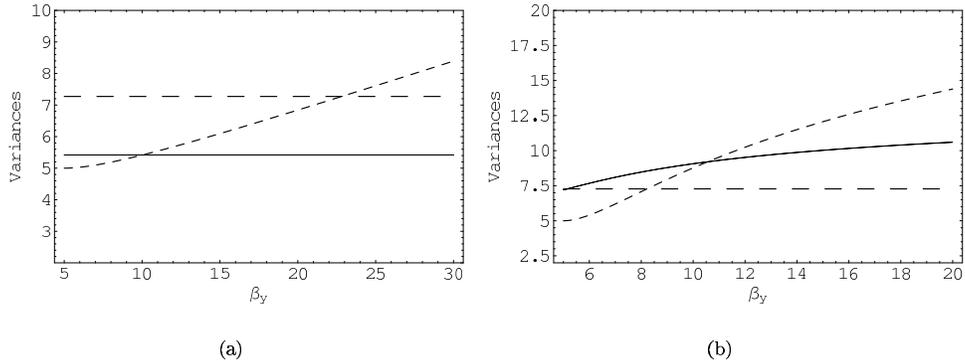}\end{center}

\caption{Variances $<\delta x^{2}>_{tot}$ (line with long dashes), $<\delta y^{2}>_{tot}$ (line with short dashes), $<\delta z^{2}>_{tot}$ (solid line) versus $\beta_{y}$ for (a) Type-1 coherent FFL and (b) Type-4 coherent FFL controlled by AND-gate. The mean protein level is fixed at m=5. The input noise is greater than that in the case of figure 2. }
\end{figure}

\section*{5. Conclusion and Outlook}

In this paper, we have studied the noise characteristics of coherent
and incoherent FFLs using the Langevin formalism as well as a numerical
simulation technique based on the Gillespie algorithm. Noise is undesirable
if it affects operational reliability. Coherent FFLs function as noise
filters and the performance of the Type-1 FFL is found to be the best
since the propagation of noise associated with the input signal is
the least in this case. The coherent Type-1 FFL is the most abundant
of FFL motifs appearing in the GTRNs of simple organisms. The functional
superiority of the Type-1 FFL, amongst the four coherent FFLs, is
the main reason why the particular motif is favoured by natural selection.
Mangan and Alon \cite{key-4} have speculated that increased effective
cooperativity of the Type-1 FFL might be responsible for its evolutionary
advantage. Thattai and van Oudenaarden \cite{key-16} have shown that
increased cooperativity leads to noise reduction. This possibly explains
why the Type-1 coherent FFL has less output noise than the other coherent
FFLs. For the incoherent FFLs, no clear conclusion regarding the role
of noise can be arrived at as concrete results are lacking. Noise
may be advantageous to function in certain cases. Stochastic resonance
is a phenomena in which noise in threshold systems facilitates detection
of subthreshold signals \cite{key-21}. In stochastic focusing, fluctuations
(noise) sharpen the response to an input signal, i.e, make a graded
response mechanism work more like a threshold one \cite{key-22}.
Further studies are needed to ascertain whether noise aids the function
of incoherent FFLs in some manner similar to stochastic focusing.
If this is true, then the most abundant motif need not be the least
noisy. Regulatory cascades of which the FFL is a special case can
exhibit interesting kinetic phenomena which include even transient
ones like pulse generation \cite{key-23,key-24}. It will be of considerable
interest to determine the effect of noise on such phenomena. 

\begin{figure}
\begin{center}\includegraphics[%
  width=4in]{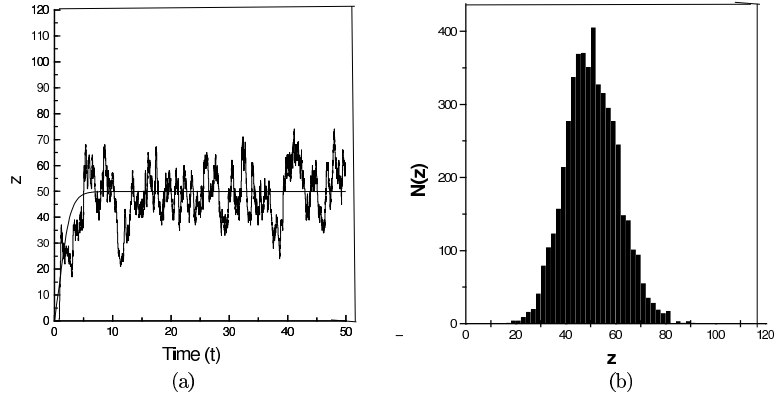}\end{center}

\caption{(a) The number of proteins z(t) as a function of time t for the Type-1 coherent FFL. The time trajectory is obtained using the GA. The solid line determines the mean curve. (b) Histogram describing the distribution of protein levels (N(z) versus z) in the steady state.}
\end{figure}

\begin{figure}
\begin{center}\includegraphics[%
  width=4in]{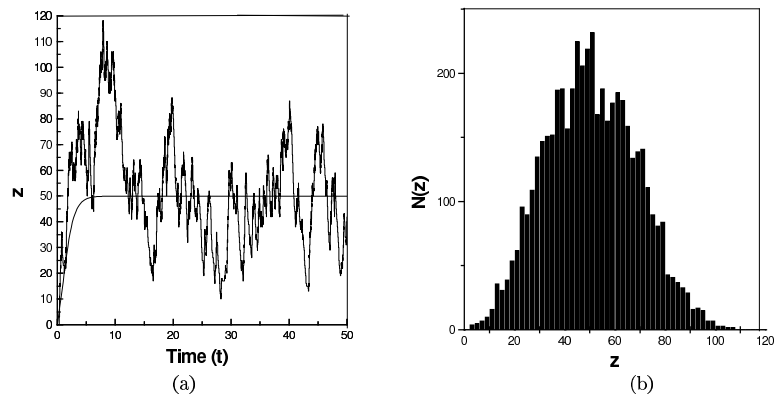}\end{center}

\caption{(a) The number of proteins z(t) as a function of time t for the Type-4
coherent FFL. The time trajectory is obtained using the GA. The solid
line determines the mean curve. (b) Histogram describing the distribution
of protein levels (N(z) versus z) in the steady state.}
\end{figure}

Fraser et al. \cite{key-14} have addressed the question of whether
noise associated with GE has any significant effect on the fitness
of an organism. They have estimated the noise in protein production
of almost all the \emph{S. cerevisiae} genes using an experimentally
verified model of stochastic GE. Their major finding is that noise
is minimized in the cases of genes for which it is likely to be most
harmful. These genes include essential genes, i.e, genes whose deletion
is lethal to the organism and genes which synthesize the subunits
of multi-protein complexes. Both types of genes are expected to be
sensitive to noise. For essential genes, fluctuations in protein levels
may have considerable effect on functional viability if the levels
fall below the threshold required for normal cellular activity. Similarly,
in the case of a multy-protein complex, fluctuations in the amounts
of protein subunits may hinder the appropriate assembly of the entire
complex. The observations of Fraser et al. are in agreement with our
results on coherent FFLs. Since noise has a deleterious effect on
the function of a coherent FFL as a persistence detector, it is minimized
in the case of the best performing Type-1 FFL.

\section*{Acknowledgements}

R.K. is supported by the Council of Scientific and Industrial Research,
India under Sanction No. 9/15 (239)/2002 - EMR-1

\noindent B.G. is supported by the Council of Scientific and Industrial
Research, India under Sanction No. 9/15 (282)/2003 - EMR-1

\end{document}